# Status of the Forward Tracker Detector of ILD

"Talk presented at the International Workshop on Future Linear Colliders (LCWS13), Tokyo, Japan, 11-15 November 2013."

Alberto Ruiz-Jimeno
(On behalf of the Forward Tracker Detector Group of the International Linear Detector (ILD) Collaboration at the International Linear Collider (ILC))

IFCA, Instituto de Física de Cantabria (CSIC-Universidad de Cantabria)
Avda. los Castros, s/n
39005 Santander-Spain

**Abstract:** An overview of the present activities and long-term R&D plans of the Spanish network for future linear accelerators aiming to design and construct the Forward Tracker Detector (FTD) system of ILD, is shown

**Introduction**

The Spanish network for future linear colliders is involved in a coordinated effort of R&D for future detectors and accelerators for particle physics. A big part of its activities is dedicated to silicon systems, aiming to precise vertex and tracking detectors. This talk intends to show the critical and beyond-the-baseline R&D activities for the construction of a FTD for the ILD experiment in the future International Linear Collider as well as the Compact Linear Collider (CLIC). After a physics-oriented motivation, a brief status report is given of the sensor design, mechanics and integration systems, data acquisition front-end electronics, powering and monitoring.

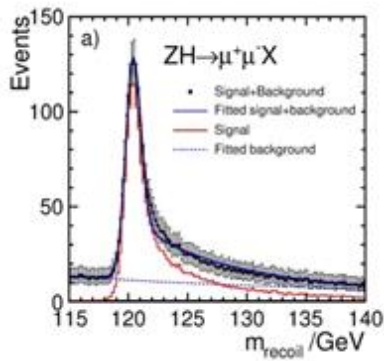

**Figure 1.** Distributions of the reconstructed Higgs recoil mass in the channel ZH →µ+µ−X (ILD Coll[3])

## The end cap tracking physics case

The low angle tracking system poses some challenges derived from the material, magnetic field and background which affect the performance of the detector. In the papers[1],[2] a detailed analysis is shown.

The goal of ILD is to achieve a good momentum resolution

$$\sigma_{1/p_T} \approx \sqrt{\left(\frac{2\times 10^{-5}}{\text{GeV}^{-1}}\right)^2 + \left(\frac{10^{-3}}{p_T[\text{GeV}]\sin\theta}\right)^2}$$

needed to do good physics analysis as e.g. to get a precise Higgs recoil mass (see Fig.1). It is also needed to have good impact parameter precision (to tag precisely heavy flavor production). In order to get the best performance ILD needs to have good pattern recognition in a full coverage angle, near to $4\pi$.

---

[1] J. Fuster et al., Forward tracking at the next e+e- collider part I: the physics case. JINST 4 P08002 (2009)
[2] S. Aplin et al. Forward tracking at the next collider Part II: experimental challenges and detector design *JINST* **8** T06001 (2013)

In some important physics analysis forward tracking good performance is increasingly important with higher center-of-mass energy. In Fig.2, as an example, it is shown the case of e+e- → $H\nu_e\nu_e$ which is particularly relevant if a heavy Higgs boson with Standard Model-like couplings is found[1].

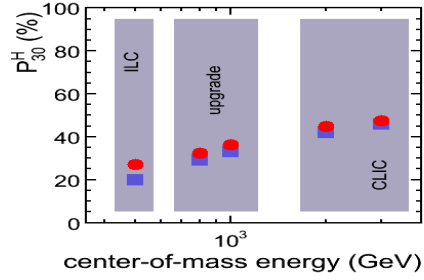

**Figure 2**. The fraction of Higgs bosons emitted in the forward direction is indicated for the W-boson fusion process (filled markers) and the Z-boson fusion process (open markers), as a function of the center-of-mass energy.

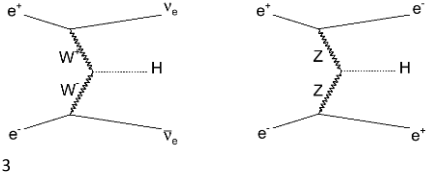

The momentum resolution for high momentum tracks is ideally given by the Gluckstern formula

$\frac{\sigma(p_T)}{p_T} = \sqrt{\frac{720}{N+4}}\, \sigma_{(r\phi)} \frac{p_T}{0.3BL^2}$    of N (>10) equally spaced layers with uniform spatial resolution $\sigma_{(r\phi)}$ in a magnetic field B (in Tesla) and with lever arm L perpendicular to the magnetic field (in meters). At ILD tracking system N is decreasing at angles below 40⁰, being less than 10 for the FTD, $\sigma_{(r\phi)}$ intrinsic is 2-6 μm for the two inner pixel disks and 7 μm for the five strip disks, and L is shorter than for the barrel tracking producing a degradation of the momentum resolution. For low momentum tracks multiple scattering contribution due to material budget is the dominant factor for the performance of the momentum resolution (see Fig.3)

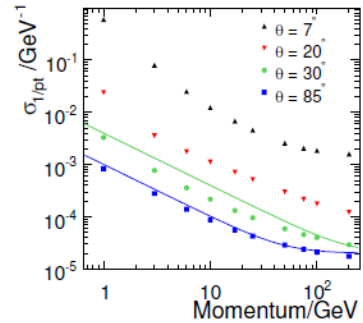

**Figure 3.** Momentum resolution as a function of the transverse momentum of particles, for tracks with different polar angles. Also shown is the theoretical expectation.

The Impact parameter resolution as a function of the momentum and polar angles is given by the expression where a= 5, b=10, for FTD[2]    $\Delta d_0 = a[\mu m] \oplus \frac{b \times \frac{L}{R}[\mu m]}{p[GeV]\cos^{3/2}\theta}.$

Beam induced background events contribute to the detector occupancy. For the innermost forward tracking disk of FTD the occupancy is simulated by e+e- → top anti-top to give by Beam-crossing (BX): $1 \times 10^{-4} \frac{hits}{mm^2} + 1.6 \times 10^{-4} \frac{hits}{BX\, mm^2}$ (average) ; $1 \times 10^{-2} \frac{hits}{mm^2} + 1.6 \times 10^{-3} \frac{hits}{BX\, mm^2}$ (peak value) which, for 10 cm long, 50 μm wide strips, gives peak occupancy of 6% per BX, well over the maximum. Therefore, higher granularity as given by pixels is required. Pixels of 25*25 mm² in the most inner region allows robust pattern recognition for a readout time of 50 μsec (about 100 BX) driving an occupancy at peak of about 10⁻⁴, which is comfortable.

The microstrip detectors in the forward tracker have radially oriented strips. To constraint the second coordinate with a low proportion of ghost hits, a stereo angle α of about 100 mrad will be used, as it is seem in Fig.4, corresponding to 100*100 mm² sensors with 25 mm pitch.

---

[3] ILC Technical Design Report: Volume 4, Part III, ILC report: ILC-REPORT-2013-040, ISBN 978-3-935702-78-2

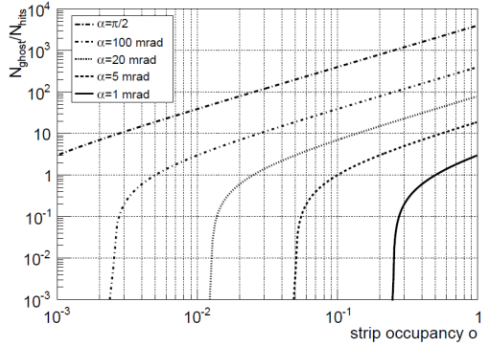

Moderately precise r-measurements should be needed in all the forward tracking layers to have a robust pattern recognition.

In the shown case, for α=100 mrad, σ(r) = 20 σ(rφ), about 100 μm for typical detectors.

$$\sigma(r\phi) = \frac{\sigma}{\sqrt{2}\cos(\alpha/2)},$$
$$\sigma(r) = \frac{\sigma}{\sqrt{2}\sin(\alpha/2)}$$

**Figure 4.** The ratio of the number of ghost hits and the number of particles traversing the sensor versus strip occupancy. The curves correspond to five values of the stereo-angle, from 1 mrad to π/2 rad. [2]

## The end cap tracker status

The end cap tracker contains seven tracking disks installed between the beam pipe and the inner field cage of the TPC. The first two are realized as pixel detectors to cope with the expected high occupancies in this area, the remaining five are strip detectors. The baseline strip sensor is a conventional microstrip sensor with integrated signal routing in a second metal layer. The baseline operational unit is the petal (composed of the sensor and standard hybrid board(s) with readout, powering and data link circuitry). In Fig.5 an scheme of the present design of FTD is shown, including the services.

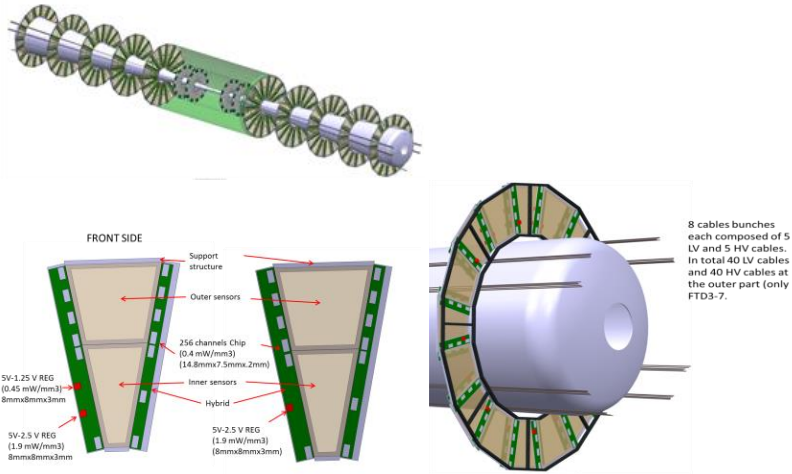

**Figure 5.** Scheme of the baseline end-cap silicon tracker of ILD

Several critical R&D activities are being done by the FTD team which are mandatory to the base-line design, including thermal management, front-end electronics and powering. At the same time there are others beyond-baseline R&D activities aiming to enhance the detector performance.

### CRITICAL BASELINE R&D

On the Si-strip detector front we aim to develop new sensor concepts based on resistive strips with signal amplification, ultralight and smart mechanics, low dissipation powering systems and low noise read out ASICs.

We are working on four R&D lines corresponding to novel pixel detectors, micro-strip detector development, thermo-mechanical studies of devices and structures for charged particle tracking and design and performance studies, aim to produce a ready-to-build engineering design of a forward tracking detector for a future linear electron-positron collider at the energy frontier (ILC or CLIC).

One of the aims of the microstrips R&D line is to enable IMB-CNM (Microelectronics National Center, Barcelona) as a qualified sensor supplier for ILD along the FTD baseline, namely: AC-coupled, 200 micron thick micro-strip sensors fabricated on six-inch wafers. IMB-CNM is an internationally recognized provider of silicon radiation sensors based on 4-inch wafers with a long experience on all available planar technology

**Thermal and Mechanical studies**

The development of ultra-transparent vertex detectors, with a very tight material budget, poses a major challenge for the thermal and mechanical design. Vertex detector ladders with a thickness of several tens of microns and a spatial resolution of well below 10 microns require very robust mechanical properties. The power generated by the sensor and ASICs must be removed with the smallest impact on the detector material.

The Spanish network institutes involved are IFCA (Institute of Physics of Cantabria, Santander) and IFIC (Institute of Corpuscular Physics, Valencia), in close collaboration with INTA (National Institute of Aerospatiale Techniques, Madrid)

The goal of the **thermal** studies is to analyze the possibility to use gas (air) as cooling, avoiding active cooling burden and reducing the material budget. That is a challenging task for inner tracking system.

There exists a partial synergy with Belle-II[4] PXD cooling system. In Fig.6 a design of the system is used. It includes small footprint fast optical Fiber Bragg Grating sensors for thermal mockup diagnostic which we are developing for use in high energy particle experiments.

A mock-up has been produced consisting of:
- stainless steel cooling blocks, enclosed with copper foil ladders and equipped with resistive heaters in the end flanges for both layers.
- a single Si thinned detector with printed Al resistors

Liquid $CO_2$ is circulating in the cooling blocks at -35 º C. $N_2$ gas cooled at 0ºC is injected towards the sensor region at 3 bar and 15L/min flow rate. The beam pipe is kept at 15ºC with a composite liquid coolant.

The measurement equipment (Fig.7), which is put inside a sealed methacrylate box, consists of:
- An infrared thermal imaging camera
- Fibber Bragg Grating (FBG) temperature and humidity sensors
- Pt110 probes

The analysis are still in progress, but are promising

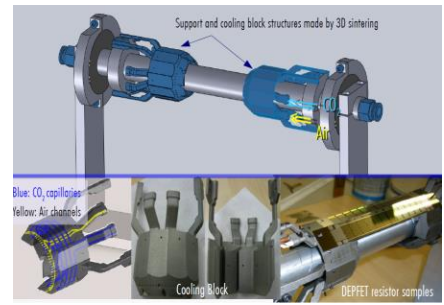

**Figure 6.** Design of the thermal system of the inner tracking system

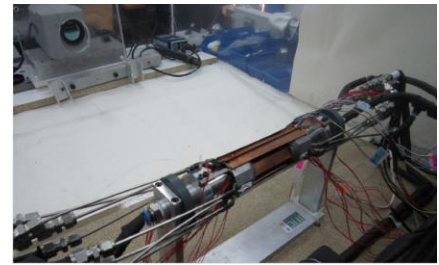

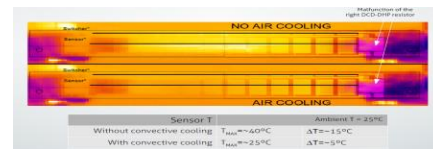

**Figure 7.** Thermal Measurement system and preliminary measurements

---

[4] **Belle II Collaboration**, T. Abe *et al.*, "Belle II Technical Design Report" arXiv:1011.0352 [physics.ins-det]. KEK-REPORT-2010-1.

**Front End Electronics**

This is a critical R&D activity. It is in an initial phase with a lot of work to be done, but with several possible fallback solutions.

The aim is to design a read-out ASIC for micro-strip detectors oriented to the ILC characteristics and timing structure. Front End readout ASICs were previously produced in the 180 and 130nm CMOS technological nodes.

Starting in 2011 in the framework of the European Project AIDA[5] (Advanced European Infrastructures for Detectors and Accelerators), a transition to ultra-deep submicron technology (65nm node) was done which yields important benefits in terms of power consumption, integration and resource sharing with other ongoing developments (Belle II, DEPFET[6] Collaboration).

The involved group is the UB (University of Barcelona) [7]

**Pulsed Powering**

During the last years the ITA group (Technological Institute of Aragon) has started a R&D line in EMC issues and Power Pulsing systems for HEP[8], in collaboration with IMB-CNM, IFCA and University of Bonn.

The front-end electronics of ILC needs to be synchronized with the small duty cycle of the beam (0.5%). Therefore, the power system (PS) has to deal with an important amount of pulsing current. Two different topologies are under study: DC-DC power converters and a solution based on super-capacitors and LV regulators. In the super-capacitors based power distribution system, the experiment will be powered remotely by small power units that supply low current to the periphery of each sub-detector. At that location, a set of super-capacitors will supply the high current pulse to the hybrids (Fig. 8).

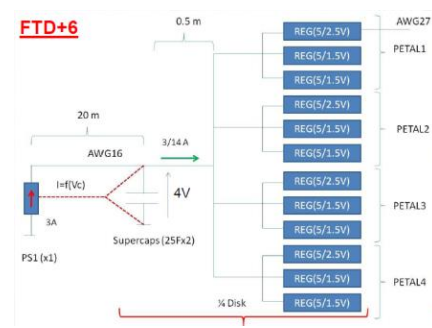

**Figure 8.** Scheme of supercapacitor based Power Supply

The high capacitance has two advantages:

– It will protects the system in case mains failure, which is similar to UPS. So it helps to shut down the system in a controlled way, depending the duration of the capacitance and voltage.

– The dynamic response of primary power unit may be very slow. Then remote regulation of the supercapacitor voltage will be easy

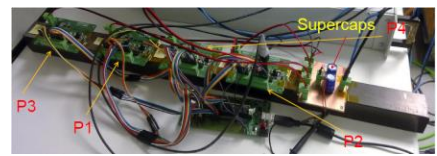

**Figure 9.** Real prototype a 1-group of FTD subdetector

There are three elements which need a deep and complete analysis to ensure the correct operation of super-capacitors for HEP applications, material budget (in fact supercapacitors are made essentially of carbon), ionizing radiation and reliability issues (under cycling operation). Radiation hardness has been analyzed and shows no stoppers, but still requires a detailed analysis.

---

[5] **AIDA** FP7 European Network, 2011-2015, Advanced European Infrastructures for Detectors at Accelerators

[6] The **DEPFET** collaboration, IEEE Trans. Nucl. Sc. 60, 2, 2 (2013)

[7] A. Dieguez, A. Montiel, R. Casanova, O. Alonso, International Workshop on Future Linear Colliders (LCWS), 2012

[8] F. Arteche et al, ECFA 2013, LCWS 2012, Texas, USA, October 2012, LC Spanish Network, Seville, Feb. 2014

DC-DC topology also presents some limitations due to reliability (millions of cycles per year) and Electromagnetic Compatibility (EMC) issues.

A real prototype of 1 FTD subdetector (Figure 9) has been developed and tested giving very good agreement with simulations. The results are very promising but a long study of the system is required in order to define the final specification

**LONG-TERM R&D**

Concerning the development of advanced tracking sensors, the team activities are based on the results from previous joint projects of CNM, IFCA and ITA groups where several microstrips technologies of interest for the ILC tracking detectors were tackled: semitransparent, thin, resistive-electrode and low gain avalanche microstrips[9]

Two main R&D lines are in progress:

- Low gain p-type segmented pixels or strips, which allow thinner sensor with same Signal/Noise. The work is based on the LGAD devices (Low Gain Avalanche Diodes) developed at previous projects

- Charge division in microstrips to reduce the complexity of double-sided sensors. The feasibility of the charge-division concept on silicon microstrip detectors to obtain two-dimensional information of the passing point of the ionizing radiation was already probed for the first time. The new R&D phase aims developing a full-fledged microstrip sensor with resistive electrodes suitable for ILC tracking requirements.

A proof-of concept prototype has been constructed (Fig.10) showing good linearity. New studies aiming to do both signal readout on the same side, using polysilicon resistive detectors, are on progress.

**CONCLUSION:**

The spanish network for Future Linear Colliders is developing a project for the end cap tracking of the International Linear Detector at the International Linear Collider and Compact Linear Collider . The teams are involved in critical R&D lines with different degree of coverage, including cooling and thermal management, pulse powering, and readout ASIC. Additional R&D aiming enhanced performance of the detectors include new detector technologies and smart mechanics.

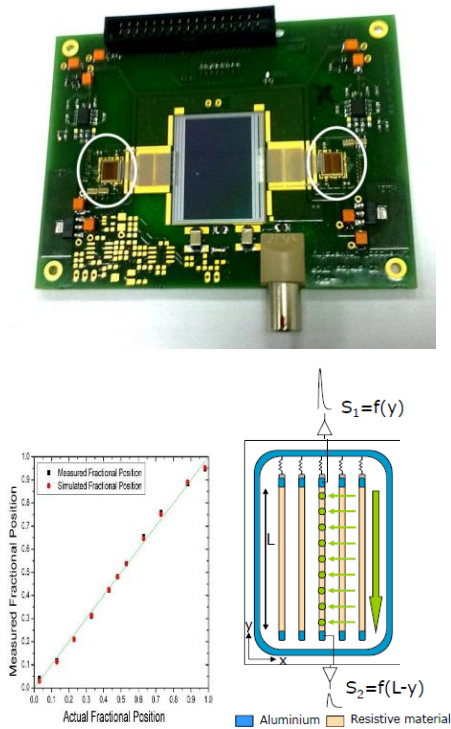

**Figure 10.** Proof-of concept prototype of charge-division in microstrip

---

[9] M. Fernández et al., Nucl. Instr. and Meth. Phys. Res. A 624 (2010) 340-343,
 D Bassignana et al., 2012 JINST 7 P02005,
 G. Pellegrini et al, RD50 funding project, 2011

The teams are moving from generic or Belle-II oriented R&D towards a coordinated and ILD oriented R&D project.